%% file: calyx-profiler.tex
\documentclass[sigplan,nonacm]{acmart} 
\setcopyright{none}
\usepackage{pervasives}

\author{Ayaka Yorihiro}
\affiliation{
  \institution{Cornell University}
  \country{USA}
}
\author{Griffin Berlstein}
\affiliation{
  \institution{Cornell University}
  \country{USA}
}
\author{Pedro Pontes García}
\affiliation{
  \institution{Cornell University}
  \country{USA}
}
\author{Kevin Laeufer}
\affiliation{
  \institution{Cornell University}
  \country{USA}
}
\author{Adrian Sampson}
\affiliation{
  \institution{Cornell University}
  \country{USA}
}

\begin{document}

\title[]{
  Understanding Accelerator Compilers\\ via Performance Profiling
}
\date{}
\hfuzz=12pt

\begin{abstract}
  Accelerator design languages (ADLs), high-level languages that compile to
  hardware units, help domain experts quickly design efficient application-specific hardware.
  ADL compilers optimize datapaths and convert software-like control flow constructs into control paths.
  Such compilers are necessarily complex and often unpredictable:
  they must bridge the wide semantic gap between high-level semantics and cycle-level schedules, and they typically
  rely on advanced heuristics to optimize circuits.
  The resulting performance can be difficult to control, requiring guesswork to find and resolve performance problems
  in the generated hardware.
  We conjecture that ADL compilers will never be perfect: some performance unpredictability is endemic to the problem they solve.

  In lieu of compiler perfection,
  we argue for \emph{compiler understanding tools} that give ADL programmers insight into how the compiler's decisions
  affect performance.
  We introduce
  \profiler, a cycle-level profiler for the Calyx intermediate language (IL).
  \profiler instruments the Calyx code with probes and then analyzes the trace from a register-transfer-level simulation.
  It maps the events in the trace back to high-level control constructs in the Calyx code to track the clock cycles
  when each construct was active.
  Using case studies, we demonstrate that \profiler's cycle-level profiles can identify performance problems in
  existing accelerator designs.
  We show that these insights can also guide developers toward optimizations that
  the compiler was unable to perform automatically,
  including a reduction by 46.9\% of total cycles for one application.
\end{abstract}

\maketitle

\section{Introduction}\label{sec:introduction}

Accelerator design languages (ADLs) are high-level languages for developing application-specific hardware accelerators.
ADL compilers must bridge a large semantic gap:
they typically start with an untimed, resource-unaware source program
and introduce cycle-level schedules,
physical resource management,
and explicit control structures such as finite-state machine (FSM) logic.

This control-logic synthesis is fundamentally complex.
ADL compilers therefore typically rely on heuristics that lead to unpredictable performance~\cite{dahlia}.
Even if the heuristics work well in most cases,
in the rare cases where they produce poor results,
they leave accelerator designers with little recourse.
The high-level ADL source code abstracts over the implementation details for the control logic,
so the only option is to inspect or simulate the generated hardware.
A waveform viewer~\cite{gtkwave, surfer} can help exhaustively visualize the generated control paths' behavior on every
cycle, but these low-level signals may have little connection to the original ADL program.
These signal traces make it difficult to understand \emph{when} a high-level action in the source program occurred or
\emph{how long} it took to run.
They can also obscure any control overheads that the compiler itself introduces, which can manifest in wasted time when
the high-level program makes no progress.


This paper argues that, because of the fundamental complexity of hardware control logic generation, ADL compilers will
never achieve perfect results for every program.
Instead, we must furnish developers with tools that help them understand how the compiler behaves.
In particular, \emph{trace-based profiling} for compiled ADL programs could help illuminate how high-level constructs
map onto cycle-level execution time.
By revealing how the compiler chose to orchestrate the events from the source program, a profiler could help developers
find and fix compiler-induced sources of inefficiency.

We present \profiler, a cycle-level profiler that differentiates
control overhead from user-defined computation. \profiler recreates a
cycle-by-cycle trace of program execution during
register-transfer-level (RTL) simulation in terms of user-defined
program blocks, and also captures timings and durations of control
events.
%
\xxx[as]{Is it maybe possible to introduce the concepts without
  referring to Calyx much? Then later we can say that we build an
  implementation using an existing intermediate language for ADL
  compilers~\cite{calyx}. This would just enhance the impression that
  our contribution is not solely tied to its embodiment in the Calyx
artifact.}  To do so, \profiler uses instrumentation to add
\emph{probes} to the original program that do not affect the cycle
count of the overall program. After obtaining a trace from RTL
simulation, \profiler uses probe signal values to reconstruct when events
were active. Events include user-defined blocks of computation and
compiler-generated control constructs. The recreated trace containing
these events is used to produce visualizations such as flame graphs
and timeline views.  \xxx[as]{Instead of ``...that users can interact
  with,'' which is sort of vague, can we get specific about what
exactly these visualizations consist of?} Our implementation of
\profiler uses the Calyx intermediate language~\cite{calyx} for ADL
compilers.

We demonstrate the usefulness of \profiler's cycle-level profiles
towards understanding the effects of existing optimization passes in the Calyx compiler (Section~\ref{sec:compiler-pass}).
\xxx[as]{I don't think readers will know what these optimization passes are. Can we instead gesture toward ``existing
  optimization passes in the Calyx compiler'' or something and use a forward-reference to the relevant section where we
discuss this?}
By using \profiler visualizations to
compare optimized and unoptimized versions of a program, users can learn about scheduling changes
and other transformations performed by a particular pass.
We also find that \profiler can help users identify performance
problems in existing designs and direct their
hand-optimization efforts to make the resulting accelerator more
performant (Section~\ref{sec:control-cycles}). Oftentimes we found
that reducing control overhead results in large performance gains, up
to 46.9\% in one application.



This paper's key contributions are:
\begin{itemize}
  \item We design \profiler, a profiler for the Calyx intermediate
    language (IL). It leverages Calyx's structure and
    instruments the program with \emph{probes} that are used to
    reconstruct the execution trace from RTL simulation.
  \item Two case studies that use \profiler to understand the effects of
    two Calyx compiler optimization passes: static promotion and
    resource sharing.
  \item Two case studies that use \profiler to find optimization
    opportunities by reducing control overhead, including a cycle-count reduction
    of 46.9\% in one application.
\end{itemize}
\xxx[ppg]{These three bullet points could be edited for parallelism
(as opposed to one full sentence and two NPs)}

\section{Example}\label{sec:example}


\begin{figure}
\begin{lstlisting}[numbers=left, belowskip=-0.8\baselineskip, escapechar=|]
X = mem;
switch X {
  case 1 { sub_one() }
  case 2 { sub_two() }
  case 3 { sub_three() } }
write_output();
\end{lstlisting}
  \caption{ADL pseudocode that uses \texttt{switch-case} to determine which subroutine to call.}
  \label{fig:switch-case-adl}
\end{figure}

\input{figures/switch-case-code}

In \cref{fig:switch-case-adl}, ADL code uses a \code{switch-case} to determine which subroutine to call.
\cref{fig:switch-case-calyx} shows the same program after compilation to Calyx.
The \code{control} block on line~\ref{switch-case:control} shows that
the \code{switch-case} was implemented using parallel composition
(lines~\ref{switch-case:par-start}--\ref{switch-case:par-end}). Each thread
runs the comparison for a specific \code{case} and the corresponding
subroutine on success. Suppose that in this
execution, \code{case} 1 passes. 
In total,
the computation in the \code{main} component during our execution will
consist of the blocks \code{read}, \code{run_s1}, and
\code{write}.

How many cycles should this execution take? Supposing each of the
computation blocks (Calyx groups) take one cycle, we may expect a
total of three cycles. However, by running this program we find
that the execution actually takes \emph{nine}. 

\begin{figure*}
  \includegraphics[width=\linewidth]{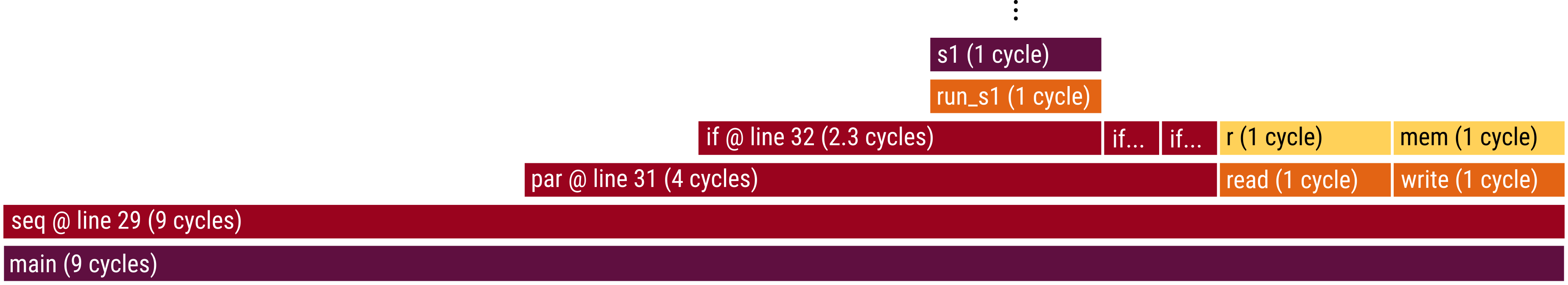}
  \caption{Flame graph from running \profiler on the Calyx program in Figure~\ref{fig:switch-case-calyx}. Elements that
    occur in parallel are normalized relative to the total length, so that summing up all threads result in the total
  length of the par. Elements called from the \code|s1| cell are omitted for space.} 
  \label{fig:switch-case-par-flamegraph}
\end{figure*}

\begin{figure*}
  \includegraphics[width=\linewidth]{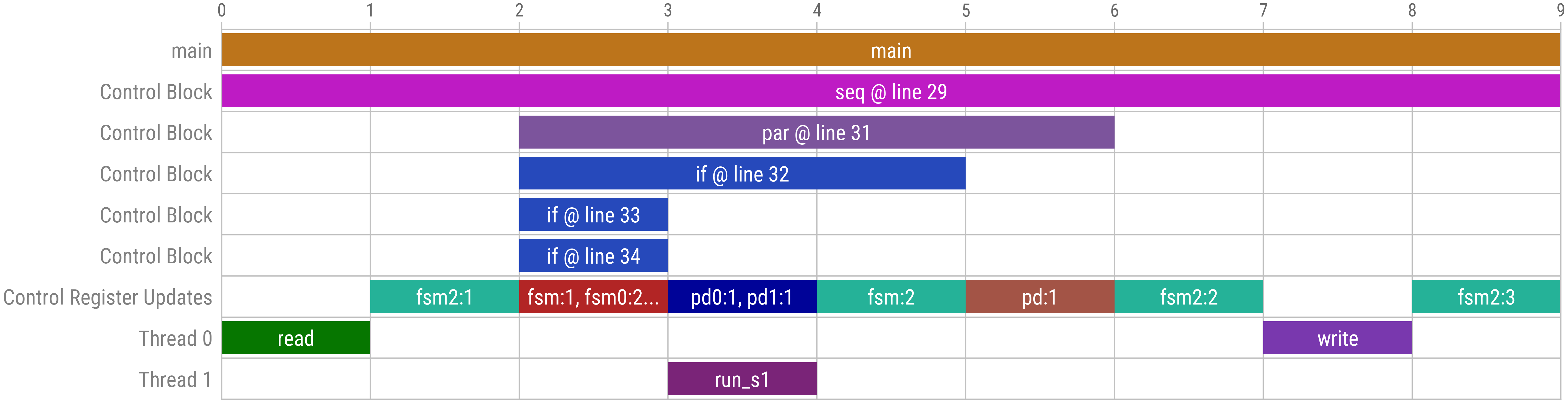}
  \caption{A timeline view from running \profiler on the Calyx program in Figure~\ref{fig:switch-case-calyx}. Elements called from the \code{s1} cell are omitted.}
  \label{fig:switch-case-par-timeline}
\end{figure*}

We diagnose this discrepancy using \profiler. \profiler produces a flame graph
(\cref{fig:switch-case-par-flamegraph}) that summarizes the
execution cycles spent on program blocks and control statements. As
expected, \code{read}, \code{run_s1}, and \code{write} each take one
cycle. But we additionally see that individual control statements (in
red) take up a total of six cycles. The gaps that exist between
control statements and program blocks are \emph{control
overhead}---cycles spent \emph{only} on control and not on
user-defined computation. This suggests that some part of our control
program introduces unexpected overhead. \xxx[g]{maybe put a single
  line about what we've learned here. Like: This suggests that
  something about our program is introducing unexpected control
processing}

To get a closer look at the control overhead, we turn to \profiler's
timeline view (Figure~\ref{fig:switch-case-par-timeline}). From the
top, the timeline view shows the whole \code{main} component (row
``main''), each control block (marked ``Control Block''),
updates to registers that manage control (marked ``Control
Register Updates''), and the two threads of user-defined program
blocks.

These control register updates are the
source of our mysterious overhead. While a user previously may have assumed
that \code{par} blocks do not incur any overhead, we see that the
compiler's structural implementation of the inner \code{if} on
line~\ref{switch-case:if} adds two control update cycles (cycle 2 and
cycle 4), and the \code{par} results in adding another control update
cycle (cycle 5). None of cycles 2, 4, and 5 contain any user-defined
computation; they exclusively manage control flow. This \code{par} overhead seems unnecessary for implementing
a \code{switch-case} since the cases are mutually exclusive.

Perhaps parallelism is not the most cycle-efficient way to implement a
\code{switch-case} because of the control overhead. We attempt a different implementation of
\code{switch-case} in Figure~\ref{fig:switch-case-calyx-fix}, where we
use a nested series of conditionals.
\begin{figure}
\begin{lstlisting}
if eq_1.out { run_s1; }
else {
  if eq_2.out { run_s2; }
  else {
    if eq_3.out { run_s3; } } }
\end{lstlisting}
  \caption{The control of a Calyx implementation of \code{switch-case}
    that uses nested \code{if-else} blocks. This control leads to a
    three-cycle reduction by eliminating the control overhead of managing
  \code{par}s.}\label{fig:switch-case-calyx-fix}
\end{figure}
Running this new version of the Calyx program, we find that it takes
six cycles for all three \code{case}s. This change eliminates the
three-cycle control overhead created by the \code{par} and the
\code{if}s within it. So, we updated the ADL compiler to replace the
\code{par} implementation of \code{switch-case} with the
\code{if-else} alternative.

\profiler provides precise cycle attribution for user-defined
computation and control overhead. The gaps in the flame graph show a
high-level view of control overhead, and the timeline gives a precise
walkthrough of \emph{when} computation and control updates
occur. While this is a relatively small example, we find in
Section~\ref{sec:control-cycles:queues:summary} that this optimization can
reduce cycle-count by 17.7\% in a larger program.
A better understanding of the compiler's behavior, through \profiler, lead directly to an optimization.

\section{Background: Calyx}\label{sec:background}


This section describes Calyx~\cite{calyx}, the intermediate language
(IL) for hardware generation that \profiler targets.

\paragraphstyle{Components} Components are encapsulations of units of hardware, similar to Verilog modules, and
analogous to functions in software languages. A component has input and output \emph{ports} to interface with other
components. Each component definition has three sections: \emph{cells} that instantiate computation elements, \emph{wires}
that connect ports, and \emph{control} that imperatively describes the component's execution
schedule. Programs begin execution at the \code{main} component.

\paragraphstyle{Cells} The \code{cells} section instantiates other components.
Each cell is an instance of either a user-defined component or a \emph{primitive} from the Calyx standard library.

\paragraphstyle{Wires and assignments} The \code{wires} section describes assignments that connect between cell ports.
Calyx assignments can contain conditional \emph{guards}, where the assignment is only active when the guard is true.

\paragraphstyle{Groups} A Calyx \emph{group} is a named, unordered set of assignments that together express a logical
operation. All assignments in the group are active simultaneously, and only when the group itself is active. For
example, the \code{read} group defined on line~\ref{switch-case:group-read} in \cref{fig:switch-case-calyx} writes the value in memory
\code{mem} to the register \code{r}. Groups in Calyx use a go-done control interface. A group is \emph{called}
when its \code{go} signal is set by another group or control, and remains active as long as the \code{go} signal is
high. In our example, the \code{read} group is called by the control block in line~\ref{switch-case:read-enable}. A
group terminates when the \code{done} condition is set; in our example, this occurs on
line~\ref{switch-case:group-read:done} to be when the register \code{r} finishes writing its new value. When this
happens, the group's \code{done} signal becomes high for one cycle.

An assignment that does not belong in a group is a \emph{continuous
assignment} (e.g., 
Figure~\ref{fig:switch-case-calyx} line~\ref{switch-case:continuous}), and is always active.

\paragraphstyle{Control}
A component's \emph{control} section is an imperative program that orchestrates the named groups.
Our example uses an \code{if} statement (ex.\ Figure~\ref{fig:switch-case-calyx}, line~\ref{switch-case:if}) to
conditionally call the \code{run_s1} group if \code{eq_1.out} is true. We also use a \code{par}
statement on line~\ref{switch-case:par-start} to parallelize the case checking and execution. Additionally, the whole
program is wrapped with the sequential composition operator, \code{seq}, at line~\ref{switch-case:seq-start}.


\xxx[ay]{Not sure where to move this to now that I'm making TDCC its own subsection}
\xxx[as]{Seems like it fits fine here? It's just more about how the language works...}
\paragraphstyle{Dynamic vs.\ Static Calyx} By default, Calyx groups are \emph{dynamically timed}, meaning that the
hardware needs to manage \code{go} and \code{done} ports. Calyx opportunistically adds \emph{static} control when the
cycle counts of groups are known, removing control overhead.
The Calyx compiler contains optimization passes to infer latencies of
dynamic groups and ``promote'' them to static whenever
possible~\cite{piezo}. In
Section~\ref{sec:compiler-pass:static-promotion}, we study the effects
of this static promotion using \profiler.

\paragraphstyle{Converting Control to Structure}

\xxx{It seems like this could be yet another LaTeX \code|paragraph| instead of a subsection? It's not really separate
from the section, in the sense that it's all about Calyx background...}

After optimization, the \tdccPass pass performs a bottom-up traversal which converts
control blocks to structural \emph{compilation groups} that manage state through registers~\cite{calyx}.
There are two kinds of compilation groups:~(1)
\tdccGroup groups, which encode sequential control (\code{seq},
\code{if}, and \code{while} statements), and \parGroup groups, which
encode \code{par} statements.

A sequential control block is encoded into a \emph{finite state
machine}, where states contain group calls (including other
control groups), condition checks (for \code{if} and \code{while}
blocks), and a special done state. The compilation group manages the
FSM state using an \code{fsm} register. Groups within
the control block are called from the \tdccGroup group with a guard
that checks the \code{fsm} register's output value. When the groups
terminate, the \code{fsm} register proceeds to the next state. The
\tdccGroup group terminates when the \code{fsm} reaches its final state.

A \parGroup group implements parallelism by calling all child groups
(including other control groups) that are inside the original
\code{par}, and finishing once all of the groups signal \code{done}
once. Since child groups may finish at different points in time, each
child group is assigned a 1-bit \parDone register for the \parGroup
group to store the child's done status. A \parDone register is set to
1 when a child group is done. The \parGroup group's \code{done}
condition is the conjunction of these \parDone registers.


\section{Trace-Based Profiling for Accelerator Designs}\label{sec:adl-profiling}

A profiler needs to provide actionable feedback to the developer by mapping raw performance data to the structure of
the source program.
When profiling software, this often corresponds to sampling the call stack and presenting execution time in terms of
which functions were active.
However, a simple call stack is not enough to represent the execution of high-level programs that compile to hardware.
Unlike software targeting CPUs, the hardware languages we seek to profile are inherently parallel.
So, we need to be able to represent this fine-grained parallelism.
We thus use the concept of call trees as an extension of call stacks and highlight the challenges that need to be
overcome to collect detailed call tree information.



We start from a cycle-by-cycle RTL simulation trace of all signals
in the circuit.  From
this we construct a \emph{\trace}, which we define as a
sequence of call trees, i.e., one call tree for every cycle.  A
\emph{call tree} is a directed graph in which nodes are cells, groups,
primitives, or control blocks. An edge exists from node \code{n1} to
node \code{n2} if \code{n1} called \code{n2}, and branches correspond to
parallelism. Call trees have special rules:
\begin{itemize}
  \item The root of any call tree is the \code{main} cell.
  \item A (non-\code{main}) cell's parents can only be groups.
  \item A control block's parents must be cells or other control blocks.
  \item A primitive node's parents must be groups.
  \item Primitive nodes are always leaves.
\end{itemize}

\begin{figure}
  \includegraphics{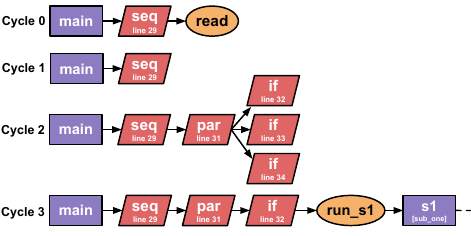}
  \caption{Call trees from cycles 0--3 of the example program in Figure~\ref{fig:switch-case-calyx}.
  Descendants of \code{s1 [sub_one]} are omitted.}
  \label{fig:activation-tree}
\end{figure}

Figure~\ref{fig:activation-tree} shows call trees of cycles 0--3 from
executing the example code in
Figure~\ref{fig:switch-case-calyx}. Rectangles represent cells from
user-defined components, parallelograms represent control blocks, and
ovals represent groups. In Cycle 2, we observe that the \code{par} on
line 31 contains three branches---one for each active thread. However,
in Cycle 3, the \code{if} statements on lines~\ref{switch-case:case2}
and~\ref{switch-case:case3} terminate, leaving the \code{par} with one
active child thread.


We use \profiler traces to produce our two main visualizations. Flame graphs
are generated by traversing the trace and identifying the number of
cycles that a unique call tree path (a call stack) appears
in. Timelines are generated by traversing the trace to identify each
node's start and end time. Paths between a cell and groups within the
corresponding component are used to organize the call hierarchy.

Software profilers can sample a call stack by examining the state of
memory at any given point in the program. In hardware, this corresponds to
observing the control signals that describe which
components are active. But FSM states and \code{go}/\code{done} port values in the
trace do not contain enough information to reproduce
accurate cycle counts of Calyx constructs
(Section~\ref{sec:implementation}). We add
information to the \emph{RTL simulation trace} using instrumentation
and analyze the source code to add information about the \emph{program structure}. These two additional sources are
enough to ``fill in the gaps'' that cannot be obtained just by looking at
the state of the control logic.

\section{Profiling Mechanism}

\begin{figure}
  \includegraphics{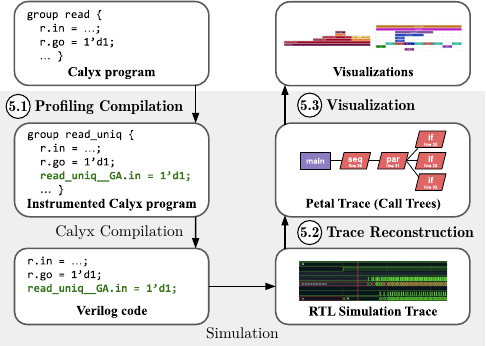}
  \caption{Overview of \profiler.}
  \label{fig:overview}
\end{figure}

Figure~\ref{fig:overview} shows \profiler's overall workflow.  There
are three phases.  The \emph{profiling compilation} phase
(Section~\ref{sec:technique:compiler}) instruments a Calyx program to
add profiling probes.  Profiling probes track group activity during
simulation without affecting timing.  The Calyx
compiler produces Verilog code, which we simulate with a standard RTL
simulator to produce a signal trace.  The \emph{trace
reconstruction} phase (Section~\ref{sec:technique:reconstruction}) constructs the \trace from
from the RTL trace and control information obtained
from the compiler (Section~\ref{sec:adl-profiling}).
The \emph{visualization} phase
(Section~\ref{sec:technique:visualizations})
produces flame graphs, the timeline view, and statistics tables from the \trace{}.
 
The source code for \profiler{} can be found on Github~\cite{ayaka}, and documentation is available online~\cite{yorihiro}.

\subsection{Profiling Compilation}\label{sec:technique:compiler}

In the profiling compilation phase, \profiler uses a special compiler flow to instrument the program with profiling
probes and record metadata about the structure of the program.

\paragraphstyle{Instrumentation}
We add an \instPass compiler pass that instruments each
group in the program with \emph{profiling probes} that track when the group
is active, and when the group calls another
group/cell/primitive. Profiling probes are implemented as wires, which are combinational; hence they do not affect the
program's cycle-level timing.

There are four types of profiling probes:
\begin{enumerate}
  \item \emph{Group active} (\code{GA}): High when a group is active. Contains the group name and the component name.
  \item \emph{Group calls group} (\code{CG}): High when a group directly calls another group. Contains the parent group
    name, the child group name, and the component name.
  \item \emph{Group calls cell} (\code{CC}): High when a group calls a non-primitive cell by via the
    \code{go} port. Contains the cell name, the group name, and the component name.
  \item \emph{Group calls primitive} (\code{CP}): High when a group calls a primitive cell. Contains the primitive cell
    name, the group name, and the component name.
\end{enumerate}
We will refer to \code{CG}, \code{CC}, \code{CP} probes collectively
as \emph{call probes}, as they encode parent-child
relationships. \code{GA} probes are active during the entire execution
of the group. Call probes are only active when the call assignment
is active.

For example, Figure~\ref{fig:probe-example} shows a group
\code{myMult} in component \code{main} instrumented with probe
assignments (bolded). The group performs multiplication using the
multiplier primitive \code{mult}, and then stores the result into the
register primitive \code{r}. The instrumentation pass sets a \code{GA}
probe high through the scope of the group
(line~\ref{probe-example:ga}). \code{mult}'s \code{go} is high as long
as its \code{done} port is not set high
(line~\ref{probe-example:mult-go}), which its corresponding \code{CP}
probe reflects (line~\ref{probe-example:cp-mult}). Similarly, \code{r}
is called when \code{mult} is done (line~\ref{probe-example:cp-r}),
which the \code{CP} probe also reflects
(line~\ref{probe-example:cp-r}).

For each component, the instrumentation pass iterates through all
groups. Within a group, the pass first creates a \code{GA} probe
assignment to high. Then, it scans all assignments within the group;
if there is an assignment to a \code{go} port of a cell, group, or
primitive, the pass creates a corresponding call probe (\code{CC},
\code{CG}, \code{CP}). The right-hand side of a call probe assignment
copies the right-hand side of the corresponding \code{go} port. Each
profiling probe is given a special annotation to prevent compiler
optimizations from removing it.

\begin{figure}
  \begin{lstlisting}[numbers=left, belowskip=-0.8\baselineskip, escapechar=|, moredelim={[is][keywordstyle]{@@}{@@}}]
group myMult {
  ... // mult primitive multiplies two numbers
  mult.go = !mult.done ? 1'd1;|\label{probe-example:mult-go}|
  r.in = mult.out;
  r.go = mult.done ? 1'd1; // r can go after mult|\label{probe-example:r-go}|
  @@myMult__main__GA.in = 1'd1;@@|\label{probe-example:ga}| // group myMult is active
  @@mult__myMult__main__CP.in = !mult.done ? 1'd1;@@|\label{probe-example:cp-mult}|
    // myMult calls primitive mult
  @@r__myMult__main__CP.in = mult.done ? 1'd1;@@|\label{probe-example:cp-r}|
    // myMult calls primitive r
  myMult[done] = r.done; } // group is done after r
  \end{lstlisting}
  \caption{A group containing multiplication instrumented with probes. Bolded lines contain probe assignments.}
  \label{fig:probe-example}
\end{figure}

\xxx[as]{Is there any way that we could provide some sort of intuition
  for \emph{why} this is the set of instrumentation probes we need?
  That is, why do we need all of them, and why don't we need any more?
Forward-references may be in order...}  Profiling probes provide
accurate cycle counts and parent-child relationships between
non-control block Calyx constructs. We know that the \code{main} cell
is active through the entire execution. Because of call tree rules
(Section~\ref{sec:adl-profiling}), groups are the only nodes that can
call cells or primitives. Thus, \code{CC} and \code{CP} probes suffice
to capture any other cell or primitive activation. Groups can be
called from other groups or directly from the cell's control. Groups
that are called from other groups are tracked with \code{CG} probes;
any groups without a corresponding \code{CG} probe (i.e., the set
  difference between groups with active \code{GA} probes and those that
are children denoted by \code{CG} probes) are called from control blocks.

\paragraphstyle{Control structure metadata}

\xxx[ay]{Tried to take a new stab at explaining this section}

To produce a faithful representation, \profiler must recreate the
structure of the original Calyx program's control prior to
optimization. To do so, we need to overcome two challenges. First, we
must distinguish between two calls of the same group at two different
points in control. Second, obtaining active durations and parent-child
relationships for control blocks requires more care as they cannot be
directly instrumented, and they ultimately get transformed into
compilation groups by the \tdccPass pass
(Section~\ref{sec:background}). We can use \code{go}/\code{done} ports of
compilation groups to find durations of compilation groups, but we
need to identify the original control block. Moreover, optimization
passes prior to the \tdccPass pass may transform control blocks.


To address both challenges, we add a compilation pass \controlPass
which runs before \instPass. To address the first challenge,
\controlPass creates a uniquely named group for each group call from
control. Then, to address the second challenge, the pass assigns a
unique identifier to each control block. We use Calyx's generic
facility for attaching arbitrary metadata to control blocks to
preserve this identifier. We modify optimization passes to preserve
this metadata through transformations. Lastly, we modify the
\tdccPass pass to use the identifier to emit a map between the created
compilation group and its original control block.

\subsection{Trace Reconstruction}\label{sec:technique:reconstruction}

\profiler uses a standard RTL simulator, such as
Verilator~\cite{verilator}, to run the instrumented program and
record a trace containing all the profiler probe signals.  In
the trace reconstruction phase, we read profiler probe signals from
the RTL simulation trace to reconstruct the \emph{\trace} consisting of call trees, which we use
to produce visualizations
(Section~\ref{sec:technique:visualizations}).

\paragraphstyle{Building the \trace}

To reconstruct the \trace, \profiler must create a call tree for every
cycle of the program by parsing the RTL simulation trace. We start by
enumerating all of the possible nodes (cells, control blocks, groups,
and primitives). The objective is to add edges that are active within
the cycle (obtained from active profiling probes and compilation
groups), which would result in a tree rooted at the vertex for the
\code{main} cell.

First, \profiler uses call probe signals to add edges between cell,
group, and primitive nodes. For example, if the \code{CP} probe
(\code{mult__myMult__main__CP}) on line~\ref{probe-example:cp-mult} of
Figure~\ref{fig:probe-example} is active, we would add an edge from
the \code{myMult} group node to the \code{mult} primitive node. If
there are any groups that are active (found from \code{GA} probe
signals) but do not have an incoming edge after adding in all edges
from call probe signals, they must be a root group of the cell. We
add edges for these groups from their corresponding cells.

Next, \profiler adds edges connecting active control block nodes into
the tree. For each active cell, we construct a \emph{tree fragment}
containing all the active control blocks within the cell---derived
from active compilation groups. We begin construction by enumerating
the cell's control blocks and selecting those that are active. Next,
we connect edges between active control blocks based on the control
program hierarchy. Finally, we insert the new fragment into the larger
tree.

\subsection{Visualization}\label{sec:technique:visualizations}

In the \emph{visualization} phase, \profiler uses \trace{}s to
produce three kinds of visualizations:~(1) flame graphs,~(2) timeline
views, and~(3) statistics tables.

The flame graph~\cite{flamegraphs} gives a summary containing total
cycles that were spent on each call tree node. Wide boxes take up a
lot of cycles and are likely bottlenecks. Boxes in the flame graph are
colored based on the call tree node type: purple is cells, red is
control blocks, orange is groups, and yellow is
primitives. \emph{Gaps} on the flame graph between the widths of an
entry box and the aggregate width of its children boxes often signify
a large control overhead.

To further gain an understanding of the ``flow'' of the program, users
can refer to timeline views. Timeline views are visualized using the
Perfetto UI~\cite{perfetto} visualization tool. Each cell gets its own
track, which contains child tracks for control blocks, control
register updates, and regular groups. Groups are assigned a thread ID
based on which (if any) \code{par} arm they appear under. Pinning
control blocks and control register updates while traversing the
timeline can help users identify root causes of gaps found in the
flame graph. To add control register updates to the timeline view,
\profiler identifies control registers in the RTL simulation trace and
tracks their signal changes.

Statistics tables help quantify bottlenecks. \emph{Cell tables}
quantify the duration of each cell activation and the control
overhead within them. \emph{Group tables} display the number of times
groups were active, and the minimum, maximum, and average duration of
the group across all calls (Section~\ref{sec:control-cycles:ffnn}).



\section{Implementation}\label{sec:implementation}

We implement \profiler in 1,075 lines of Rust
and 3,720 lines of Python.
\emph{Trace reconstruction} uses the \code{vcdvcd}~\cite{vcdvcd} library to parse
and process VCD traces, and \emph{visualization} uses
Perfetto~\cite{perfetto} and Gregg's original SVG-based flame graph
visualization tool~\cite{flamegraphs}.

\xxx[as]{One simple thing that implementation sections sometimes have here, at the top, is a brief summary of what we
  did to implement the idea. For example: ```We implement \profiler in [number] lines of Python. Trace reconstruction
  uses [library] to parse and process VCD traces, and visualization uses Perfetto and Gregg's original SVG-based flame
graph visualization tool.'' Something at that level of detail?}

\paragraphstyle{Alternative profiling techniques}
\profiler uses instrumentation to trace executions.
At first, we attempted to avoid instrumentation by directly interpreting
FSM control register values using a map between states in an FSM and
the groups that they represented. However, this approach failed to distinguish control-overhead cycles from actual
group execution cycles.
Next, we explored using
Calyx's \code{go}/\code{done} control signals to track active groups
(Section~\ref{sec:background}).
While this approach accurately measured groups' actual
cycle counts, not all user-defined groups persist through optimization
passes. Also, there is no straightforward way to use \code{go}/\code{done}
signals to track parent-child relationships between groups that call
other groups. Instrumentation addresses both of these problems.

\paragraphstyle{Instrumentation} 
We needed to ensure that probes did not affect the program's cycle-level
timing.
One challenge involved primitive cells \code{p} whose
\code{CP} assignments contained the negation of \code{p}'s done port
(e.g., line~\ref{probe-example:cp-mult} in
  Figure~\ref{fig:probe-example} uses \code{!mult.done} as the
guard). During compilation, a pass divided the original group such
that the probe assignment only ran during the execution of the
primitive. The \code{p.done} in the probe guard was thus redundant,
but it interfered with an existing resource sharing optimization pass. To resolve this, we
added a pass that would replace \code{p.done} in this pattern with a constant false signal.
%
Additionally, we prevented optimization passes from removing probes
and their assignments by implementing a \code{@protected} attribute
that prevents Calyx passes from removing otherwise-dead cells and assignments.

\section{Case Studies: Understanding Compiler Choices}\label{sec:compiler-pass}

ADL compilers are equipped with optimization passes to improve the
resulting hardware's performance. However, these passes rely on
heuristics and have to optimize multiple performance metrics (cycle
counts, area, and frequency). So, they may not yield optimal
performance. Previously, there was no straightforward way for users to
understand the effects of transformations made by optimization
passes. \profiler visualizations aid users by showing precise group
execution times and control overheads, which we demonstrate through
case studies of two compiler passes:
Section~\ref{sec:compiler-pass:static-promotion} studies how the
Static Promotion pass~\cite{piezo} optimizes cycle counts, and
Section~\ref{sec:compiler-pass:cell-share} studies how the Resource
Sharing pass~\cite{calyx} inhibits cycle count
optimizations in favor of optimizing area usage.

\subsection{Static Promotion}\label{sec:compiler-pass:static-promotion}

The \emph{static promotion} optimization pass~\cite{piezo} converts
originally dynamic groups into \code{static} groups to avoid control register
update cycles.
%
We use the Dahlia implementation of the \code{linear-algebra-2mm}
Polybench~\cite{polybench} benchmark program compiled to Calyx.

\begin{figure}
\begin{lstlisting}[numbers=left, belowskip=-0.8\baselineskip, escapechar=|, moredelim={[is][keywordstyle]{@@}{@@}}]
seq {
  read_A_idx; // store value in reg_a
  read_B_idx; // store value in reg_b
  mult_A_B; // multiply reg_a and reg_b
  write; // write to memory
  i_next; // initializes value for loop
  ... } // loop
\end{lstlisting}
  \caption{\code{control} block snippet that multiplies indices from two memories.}
  \label{fig:compiler-pass:static-promotion:code}
\end{figure}

Specifically, we focus on the simplified code snippet shown in
Figure~\ref{fig:compiler-pass:static-promotion:code}. The
\code{read_A_idx} group reads an index of memory \code{A} and stores
the value into register \code{reg_a}. Similarly, the \code{read_B_idx}
group reads an index of memory \code{B} and stores the value into
register \code{reg_b}. The \code{mult_A_B} group performs \code{reg_a}
$\times$ \code{reg_b}, and stores the result into a register
\code{res}. \code{write} writes the value of \code{res} into memory
\code{B}. Lastly, \code{i_next} initializes a value for the next steps
of computation, containing a loop.

\begin{figure}
  \centering
  \makebox[0.45\textwidth][c]{
    \begin{subfigure}{0.49\textwidth}
      \includegraphics[width=\linewidth]{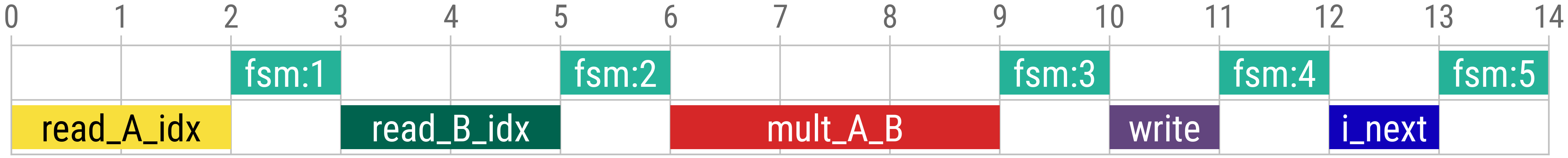}
      \caption{Timeline view \emph{without} static promotion (14 cycles).}
      \label{fig:static-promotion:disabled}
  \end{subfigure}}
  \\[2mm]

  \makebox[0.45\textwidth][c]{
    \begin{subfigure}{0.49\textwidth}
      \includegraphics[width=\linewidth]{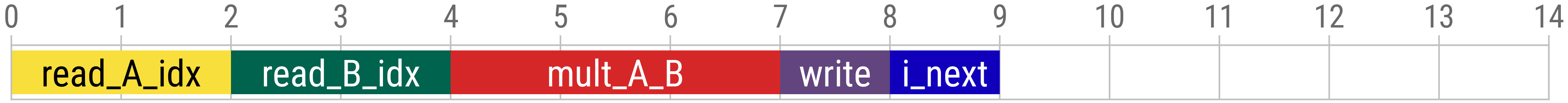}
      \caption{\textit{Expected} timeline view with static promotion (9 cycles).}
      \label{fig:static-promotion:expected}
  \end{subfigure}}
  \\[2mm]

  \makebox[0.45\textwidth][c]{
    \begin{subfigure}{0.49\textwidth}
      \includegraphics[width=\linewidth]{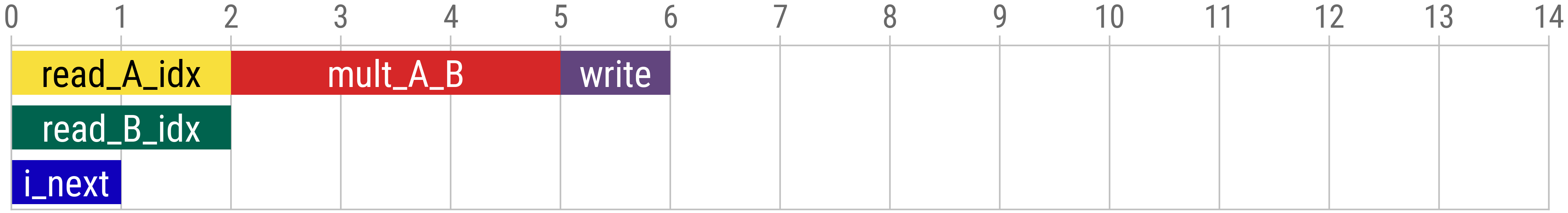}
      \caption{\textit{Actual} timeline view with static promotion (6 cycles).}
      \label{fig:static-promotion:enabled}
  \end{subfigure}}
  \caption{Timeline views of the code snippet in Figure~\ref{fig:compiler-pass:static-promotion:code} with and without
  static promotion optimization.} 
  \label{fig:compiler-pass:static-promotion:timeline}
\end{figure}

Figure~\ref{fig:static-promotion:disabled} shows the timeline view
snippet of this code snippet \emph{without} static promotion, which
takes 14 cycles. In the top track, we see updates to the FSM that
manage control in the snippet. The order of events is exactly as the
program states, with an FSM register update after every group's
execution. We find that we have a five-cycle control overhead resulting
from those FSM register updates.

From a high-level understanding of the static promotion pass, a user
may expect that running the static promotion pass for this snippet
will straightforwardly lead to a nine-cycle execution. The expected
timeline view would look like
Figure~\ref{fig:static-promotion:expected}, where groups occur
immediately after another, and the five-cycle control overhead is
eliminated. However, when the user runs the program through static
promotion, they are surprised to find that the snippet only took them
\emph{six} cycles---a much larger optimization than expected!

How did we get a more optimized execution? We can
turn to the timeline view shown on
Figure~\ref{fig:static-promotion:enabled}. First, we notice that there
is no control register update track at the top, since static promotion
was able to convert the whole snippet into \code{static}. We also
observe that the compiler rescheduled groups \code{read_A_idx},
\code{read_B_idx}, and \code{i_next} to run in parallel, since it
inferred that there were no dependencies between the three groups. By
using \profiler, the user can now understand scheduling effects of the
static promotion compiler optimization, including parallelism between
groups that they did not intentionally include.


\subsection{Resource Sharing}\label{sec:compiler-pass:cell-share}

Resource sharing is an area resource optimization that reuses existing
circuits to perform disjoint computations. For example, if two
registers were used sequentially,
the compiler could use one register to run the functionality of
both. Saving resources is important as FPGAs have limited resources,
and the cost of building ASICs depends on the necessary
resources. However, by using a single resource, resource sharing may
\emph{disable} other cycle count optimizations.

Similar to Section~\ref{sec:compiler-pass:static-promotion}, we study
a Calyx implementation of the \code{3mm} benchmark from
Polybench~\cite{polybench}. The \code{3mm} program performs matrix
multiplication between matrix pairs and then multiplies the results
together, yielding $(A \times B) \times (C \times D)$.

We run the program under the standard Calyx optimization passes with
resource sharing enabled, which takes 14,259 cycles. However, with
resource sharing disabled, we get 8,481 cycles. Resource sharing \emph{adds} a $1.7\times$ cycle count overhead!

\begin{figure}
  \centering
  \begin{subfigure}{0.45\textwidth}
    \includegraphics[width=\linewidth]{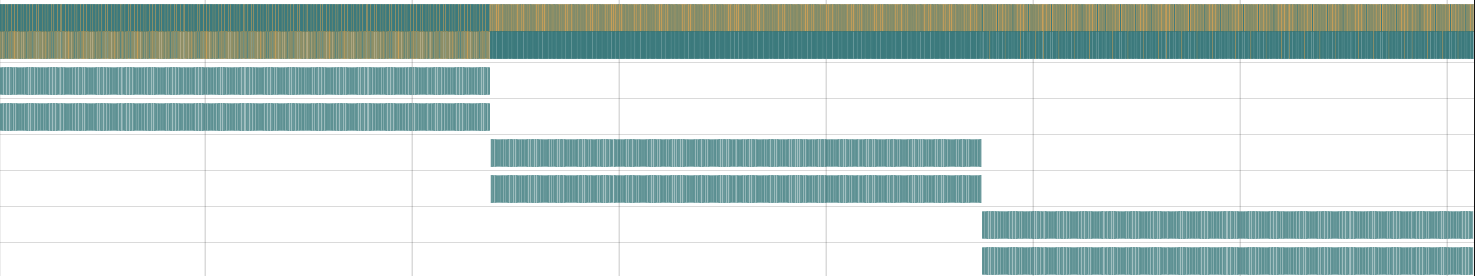}
    \caption{Resource sharing is \emph{enabled} (14,259 cycles).}
    \label{fig:cell-share:timeline:enabled}
  \end{subfigure}
  \vspace{1mm}
  \begin{subfigure}{0.45\textwidth}
    \includegraphics[width=\linewidth]{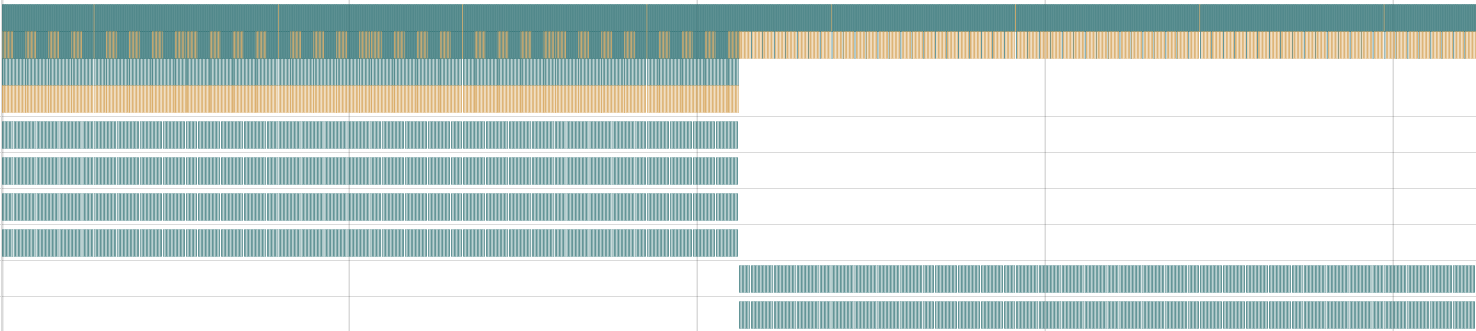}
    \caption{Resource sharing is \emph{disabled} (8,481 cycles).}
    \label{fig:cell-share:timeline:disabled}
  \end{subfigure}
  \caption{Full-program timeline views of the full \code{3mm} program execution, with and without resource sharing.}
  \label{fig:cell-share:timeline}
\end{figure}

Running \profiler on both versions of the program produces the
timeline views in Figure~\ref{fig:cell-share:timeline}. We can observe that the resource-sharing-disabled version of the
program contains more parallelism. Additionally, the
resource-sharing-enabled version of the program has three ``sections''
corresponding to the three matrix multiplication blocks, whereas the
resource-sharing-disabled version has only two. A detailed
investigation shows that in the resource-sharing-disabled version, the
two initial multiplications (\code{A} $\times$ \code{B} and \code{C}
$\times$ \code{D}) are run in parallel. As a result, we can conclude
that resource sharing shared cells between the two initial
multiplications which forced them to happen in
sequence when they could be done in parallel.

At this point, it is up to the user to determine which of area or
cycle counts they would like to save more, or for them to derive a
more optimized version of the program that saves both. However,
\profiler helps the user understand attribution of cycle counts under
this tradeoff.

\section{Case Studies: Optimizing User Programs}\label{sec:control-cycles}

\xxx[as]{I recommend a title change (and tweak to the first couple of sentences) here that tries to focus more
  specifically on control overhead.
  Echoing a theme from the intro: the idea is that a main thing that an ADL compiler does for you is synthesize control logic.
  When this is less than optimal, it can cause problems that look like ``dark matter'' from the programmer's perspective.
\profiler helps you see that otherwise-invisible cost.}

\profiler helps users identify and fix performance bottlenecks,
especially ones that do not directly come from user-defined
computation. In the process of synthesizing control logic, ADL
compilers may introduce control overhead in unexpected and
``invisible'' ways to the user. \profiler demystifies this nebulous
cost. The flame graph is useful for identifying high-level
discrepancies between expected and actual bottlenecks, timeline views
give a detailed trace, and statistics tables quantify targets for
optimization. We demonstrate this through two case studies:
Section~\ref{sec:control-cycles:ffnn} studies optimizations we made on
a forward-feeding neural network program;
Section~\ref{sec:control-cycles:queues} walks through optimization
strategies applied to a 1.4-million-cycle packet-scheduling hardware
design. 


\subsection{Forward Feeding Neural Network}\label{sec:control-cycles:ffnn}

\paragraphstyle{Setup} For our first case study, we evaluate a Calyx
implementation of a forward feeding neural network (FFNN) model,
produced by compiling a PyTorch model via Allo~\cite{allo} to
Calyx. The FFNN model has an input of 64 features. It contains a fully
connected layer of size $64 \times 48$, a ReLU activation, and a
second fully connected layer of size $48 \times 4$. Our initial
implementation takes 409,715 cycles.
\begin{figure}
  \makebox[0.45\textwidth][c]{
    \begin{subfigure}{0.49\textwidth}
      \includegraphics[width=\linewidth]{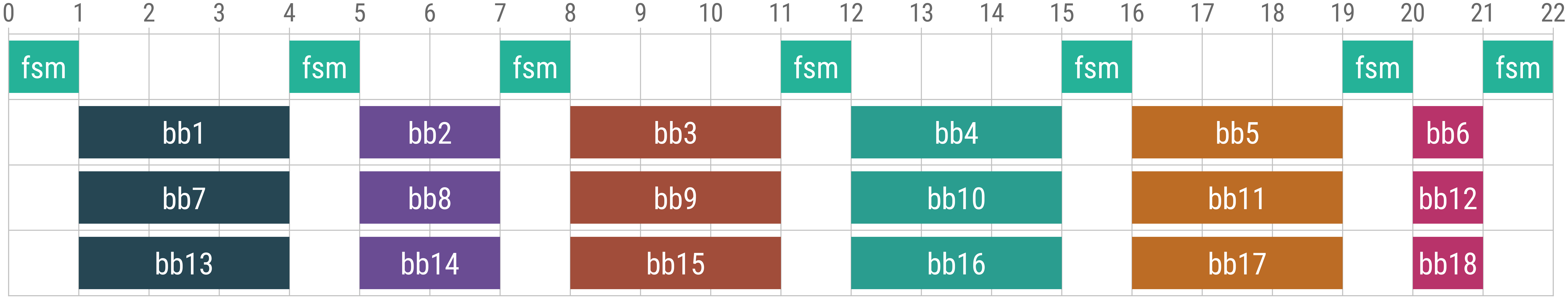}
      \caption{Original program: Total 22 cycles.}
      \label{fig:control-cycles:ffnn-timeline:original}
  \end{subfigure}}
  \vspace{2mm}
  \makebox[0.45\textwidth][c]{
    \begin{subfigure}{0.49\textwidth}
      \includegraphics[width=\linewidth]{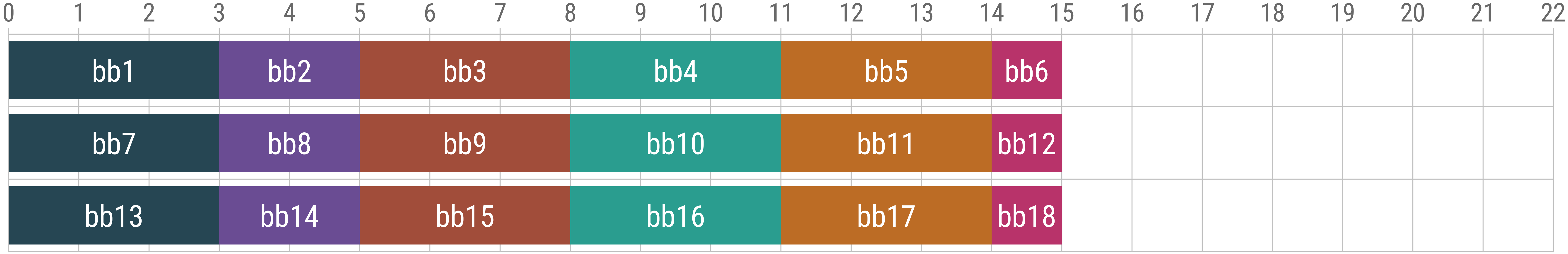}
      \caption{Program optimized with \texttt{static}: Total 15 cycles.}
      \label{fig:control-cycles:ffnn-timeline:optimized}
  \end{subfigure}}
  \caption{Timeline views of two versions of a snippet of the FFNN program.}
  \label{fig:control-cycles:ffnn-timeline}
\end{figure}

\paragraphstyle{Identifying the problem}
Using the profiler's cell overview statistics, we find that only 62\%
of these cycles are user-defined computation cycles, which presents to
us an opportunity for optimization. This program has the
\code{forward} component do all of the computation, and since the
\code{main} component only exists as a wrapper, we know that
\code{forward}is the component to optimize. To take a closer look at
a potential control bottleneck, we refer to the timeline view;
Figure~\ref{fig:control-cycles:ffnn-timeline:original} shows a close-up
snippet of the timeline. In this snippet, we can see that there are
multiple parallel sequences running in lockstep (for example, both
\code{bb_1} and \code{bb_6} run for three cycles), with an FSM update
cycle in between (one FSM for each \code{par} track, but all updating
simultaneously). This sequence occurs in a loop in the program
execution, and we confirm that the cycles where FSM updates happen do
not have any other activity, marking it as control overhead.

The timeline revealed to us that the \code{control} program of our
snippet uses dynamic scheduling. This control overhead is necessary
because we and the compiler are not certain that groups in the snippet
will have a fixed latency. However, if we can deduce that groups will
always have a fixed latency, we can eliminate the control overhead by
switching from dynamic to static code.


\xxx[as]{``Static promotion'' needs a back-reference, citation, or
  otherwise further explanation so readers know what it is. Also, we
  need a little intuition here about why that would solve the
  control-overhead problem. We have said there are wasted cycles, but
  the reader needs to be told that (a) this code is using dynamic
  scheduling, which comes with that overhead inherently for a good
  reason, and (b) one way to solve these wasted cycles is to switch
from dynamic to static code.}

Before we refer to code to figure out
whether we can convert groups to static, we can obtain evidence from
\profiler's group statistics table
(Section~\ref{sec:technique:visualizations}).
Table~\ref{fig:control-cycles:ffnn:groups-table} shows
group statistics for groups \code{bb_0} to \code{bb_6}. The ``Min'',
``Max'', and ``Avg'' columns show the minimum, maximum, and average
number of cycles for which the group was active. The ``Times
Active'' column shows the number of times the group was active, and
the ``Total'' cycle shows the total number of cycles for which the
group was active. Since the ``Min'', ``Max'', and ``Avg'' columns all
have the same value for each row, we gain greater confidence that
these groups can be converted to static, and we can eliminate the
control register update cycles in between them as a result.

\paragraphstyle{Solution} After confirming that each group will only
be active for the stated ``Avg'' number of cycles, we annotate all of
the groups of interest as static code. The optimized version of the
program contained 287,603 cycles, yielding a \ffnnOptPer cycle count
cut. Referring to the timeline view for this optimized execution
(Figure~\ref{fig:control-cycles:ffnn-timeline:optimized}), we find
that this optimization had the expected effect of removing the control
overhead we previously found in
Figure~\ref{fig:control-cycles:ffnn-timeline:original}.

\begin{table}
  \begin{center}
    \begin{tabular}{ lrrrrr }
      \toprule
      \textbf{Group} & \textbf{Min} & \textbf{Max} & \textbf{Avg} & \textbf{Times Active} & \textbf{Total} \\
      \midrule
      \texttt{bb\_1} & 3 & 3 & 3 & 9,216 & 27,648 \\
      \texttt{bb\_2} & 2 & 2 & 2 & 9,216 & 18,432 \\
      \texttt{bb\_3} & 3 & 3 & 3 & 9,216 & 27,648 \\
      \texttt{bb\_4} & 3 & 3 & 3 & 9,216 & 27,648 \\
      \texttt{bb\_5} & 3 & 3 & 3 & 9,216 & 27,648 \\
      \texttt{bb\_6} & 1 & 1 & 1 & 9,216 & 9,216 \\
      \bottomrule
    \end{tabular}
  \end{center}
  \caption{Group statistics obtained from the initial FFNN program. \code{bb_7-12}, and \code{bb_13-18} follow.}
  \label{fig:control-cycles:ffnn:groups-table}
\end{table}

\subsection{Packet Scheduling Queues}\label{sec:control-cycles:queues}

\xxx[as]{A more descriptive title for this would say more about the application: e.g., ``Packet Scheduler'' or ``Packet
Scheduling Queues.''}

\xxx{Fix this description using descriptions similar to Piezo paper}
\paragraphstyle{Setup} In our second case study, we profile a PIFO
tree packet scheduler created by the Python-Calyx packet scheduler
generator~\cite{piezo}. The PIFO tree decides allocations of bandwidth
between multiple network flows. Specifically, the program we profile
implements a \emph{strict} policy with six input flows managed by FIFO
queues, containing a total of 20,000 commands (between push and
pop). The program contains a \code{pifo} component that manages the
six FIFOs to perform push and pop commands.  The program originally
took 1,489,329 cycles.

In this case study, we identify multiple causes of control
overhead. Section~\ref{sec:control-cycles:queues:switch-case}
discusses how we found the parallel \code{switch-case} discussed in
Section~\ref{sec:example}; Section~\ref{sec:control-cycles:while}
demonstrates how we used \profiler to identify a compiler performance
bug in Calyx \code{while} loops conditions;
Section~\ref{sec:control-cycles:queues:summary} summarizes how our
control overhead optimization techniques led to a 698,129 cycle
optimized version (an overall~\strictOptPer cycle reduction) of the
packet scheduling program.

\subsubsection{\code{switch-case} Implementation}\label{sec:control-cycles:queues:switch-case}

\begin{figure}
  \includegraphics[width=\linewidth]{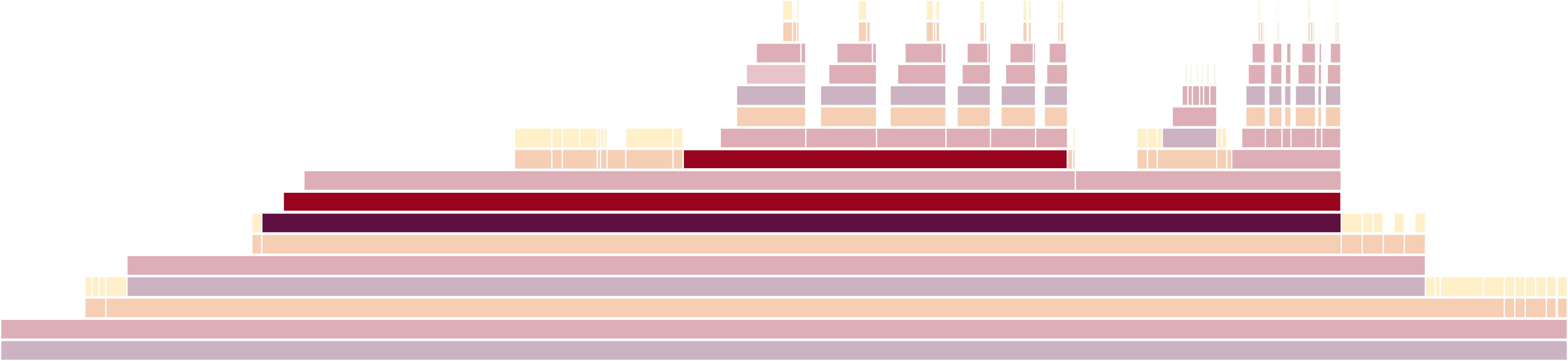}
  \caption{Flame graph of the packet scheduling program.}
  \label{fig:control-cycles:queues-flamegraph}
\end{figure}

\paragraphstyle{Identifying the problem} The profiler produces the
flame graph on Figure~\ref{fig:control-cycles:queues-flamegraph}. From
the lengths of the red boxes (representing control blocks), we observe
that control blocks take up a significant amount of time in the
execution. Furthermore, it seems that the \code{myqueue} cell (the
bottom-most, purple block) which instantiates the \code{pifo}
component takes up a significant portion of the total cycles spent. It
also devotes a large percentage of its cycles to control register
updates, as indicated by the large horizontal gaps. The two bolded
control node blocks are \code{par} blocks, which suggests that the
\code{pifo} component involves nested parallelism. We decide to
attempt optimizing the \code{pifo} component.

\begin{figure}
  \includegraphics[width=\linewidth]{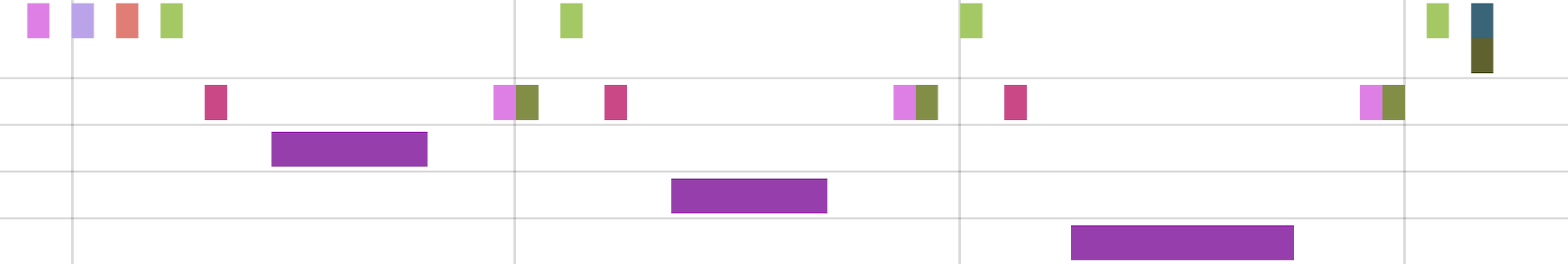}
  \caption{Timeline of groups active during one call of the \code{myqueue} cell.}
  \label{fig:control-cycles:queues:pifo-timeline}
\end{figure}

To take a closer look, we refer to the timeline view in
Figure~\ref{fig:control-cycles:queues:pifo-timeline}, which shows us
the groups that were active in one call of \code{myqueue}. The shape
of the timeline shows that, despite the program's nested parallelism,
the groups actually run
\emph{sequentially} (except for the compiler-optimized parallelism on the
first row). Scheduling parallelism incurs overhead via \parDone
register updates, so there may be an optimization opportunity in the
nested \code{par} blocks.  By inspecting the Python-Calyx packet
scheduler source code, we find that these \code{par} blocks come
from the implementation of \code{switch-case} in the Python-Calyx
generator. We thus find that the \code{switch-case} implementation in the
Python-Calyx generator adds unnecessary control register update
cycles, as described in Section~\ref{sec:example}.


\paragraphstyle{Solution} As described in Section~\ref{sec:example},
we changed the implementation of \code{switch-case} in the Python-Calyx
generator to use nested \code{if} statements instead of parallelism.
The resulting program took 1,224,589 cycles because \code{switch-case}
was used in multiple cells. Thus, a small change inspired
by the profiler made an overall 17.7\% cut on cycle counts.

\subsubsection{While Loop Conditions}\label{sec:control-cycles:while}

In this subsection, we describe a code pattern (\code{while} loop
conditions) exhibiting unnecessary control overhead that occurred in
multiple contexts in the packet scheduling program. We present a
simplified example of this pattern.


\xxx[as]{Here's a suggestion for reorganizing this subsection to make it easier to digest. (This may apply to other,
  similar case-study subsections as well.) Break it into three chunks, possibly delimited by LaTeX \code|paragraph|s:
  setup, finding the problem, and implementing a solution.
  Critically, the ``setup'' chunk would just describe the program, not anything about how it's compiled.
  Then, ``finding the problem'' would tell the story of how the profiler helped elucidate some kind of performance
  problem and its root cause.
  In this case, Figure~\ref{fig:control-cycles:while:while-desugar-ex} belongs in this ``finding the problem'' chunk
  because it helps explain \emph{why} we see the extra control overheads: it is the root cause of the phenomenon that
  the profiler revealed.
Finally, the ``solution'' chunk can talk about what concrete action we take in response to understanding the problem.}


\code{while} loop semantics in Calyx allow for the \code{with} keyword
with a combinational group that computes the
condition. Figure~\ref{fig:control-cycles:while:while-comb-ex} shows a
\code{while} loop using a combinational group \code{comb}, which
contains assignments to the \code{cond} cell.

The \code{with} keyword and combinational group is implemented by the
compiler as Figure~\ref{fig:control-cycles:while:while-desugar-ex}.
Here, the \code{cond_group}group (non-combinational) computes the
condition and writes to the register
\code{cond_reg}. \code{cond_group} is called both before the
\code{while} loop, and in the last step of the body.

\begin{figure}
  \begin{subfigure}{0.45\textwidth}
  \begin{lstlisting}[numbers=left, belowskip=-0.8\baselineskip, escapechar=|]
while cond.out with comb {
  seq {
    ... // body
  }
}
  \end{lstlisting}
    \vspace{2mm}
    \caption{\texttt{while} loop using \texttt{with} combinational group.}
    \label{fig:control-cycles:while:while-comb-ex}
    \vspace{2mm}
  \end{subfigure}
  \begin{subfigure}{0.45\textwidth}
    \begin{lstlisting}[numbers=left, belowskip=-0.8\baselineskip, escapechar=|]
cond_group; // write cond result to cond_reg
while cond_reg.out {
  seq {
    ... // body
    cond_group; // write cond result to cond_reg
  }
}
    \end{lstlisting}
    \vspace{2mm}
    \caption{Compiler implementation of the \texttt{while} block.}
    \label{fig:control-cycles:while:while-desugar-ex}
    \vspace{2mm}
  \end{subfigure}
  \caption{Two versions of \texttt{while} loops in Calyx.}
  \label{fig:control-cycle:while:ex}
\end{figure}

\begin{figure}
  \begin{subfigure}{0.45\textwidth}
  \begin{lstlisting}[numbers=left, belowskip=-0.8\baselineskip, escapechar=|]
seq {
  init; // initialize counter
  while lt.out with cond { // counter == bound? |\label{fig:control-cycles:while:cond-group}|
    par {
      seq { read; update; write; }
      incr; // increment counter
    } } }
  \end{lstlisting}
    \vspace{2mm}
    \caption{Original \code{while} using a combinational group.}
    \vspace{2mm}
    \label{fig:control-cycles:while:code:original}
  \end{subfigure}

  \begin{subfigure}{0.45\textwidth}
  \begin{lstlisting}[numbers=left, belowskip=-0.8\baselineskip, escapechar=|]
seq {
  init; // initialize counter
  cond_group; // write cond result to cond_reg
  while cond_reg.out { |\label{fig:control-cycles:while:cond-group}|
    seq {
      par {
        seq { read; update; write; }
        incr; // increment counter
      }
      cond_group;
    } } }
  \end{lstlisting}
    \vspace{2mm}
    \caption{\code{while} without combinational group.}
    \vspace{2mm}
    \label{fig:control-cycles:while:code:desugared}
  \end{subfigure}

  \begin{subfigure}{0.45\textwidth}
  \begin{lstlisting}[numbers=left, belowskip=-0.8\baselineskip, escapechar=|]
seq {
  init; // initialize counter
  cond_group; // write cond result to cond_reg
  while cond_reg.out { |\label{fig:control-cycles:while:cond-group}|
      par {
        seq { read; update; write; }
        seq { incr; cond_group; }
          // increment and check counter
      } } }
  \end{lstlisting}
    \vspace{2mm}
    \caption{Parallel optimization of incrementing and checking the group.}
    \vspace{2mm}
    \label{fig:control-cycles:while:code:smart}
  \end{subfigure}
  \caption{Example \code{control} programs of \texttt{while} with conditions pattern and optimizations.}
\end{figure}
\begin{figure}
  \centering
  \makebox[0.45\textwidth][c]{
    \begin{subfigure}{0.49\textwidth}
      \includegraphics[width=\linewidth]{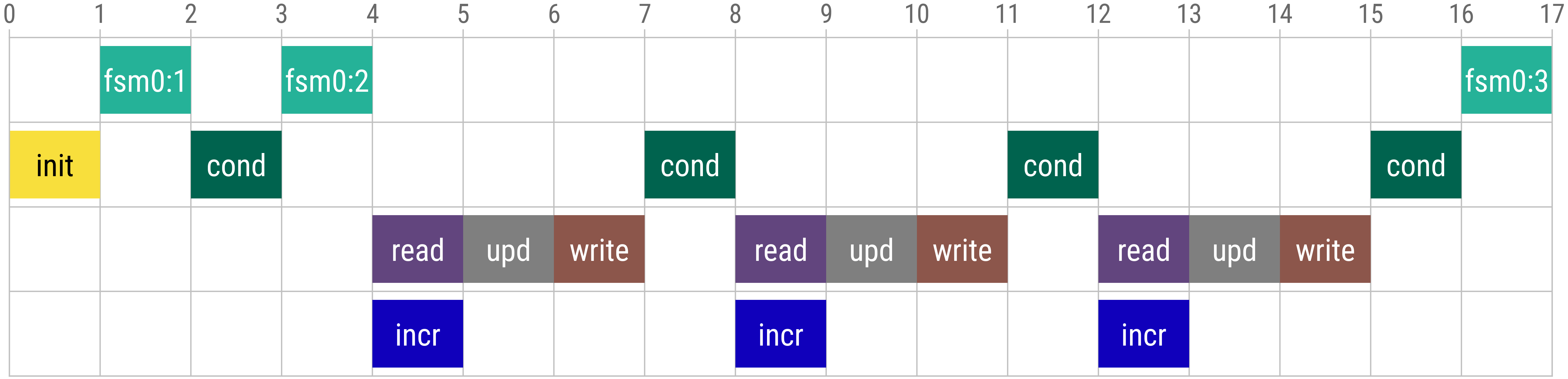}
      \caption{Original program: Total 17 cycles.}
      \label{fig:control-cycles:while:original-timeline}
  \end{subfigure}}

  \makebox[0.45\textwidth][c]{
    \begin{subfigure}{0.49\textwidth}
      \includegraphics[width=\linewidth]{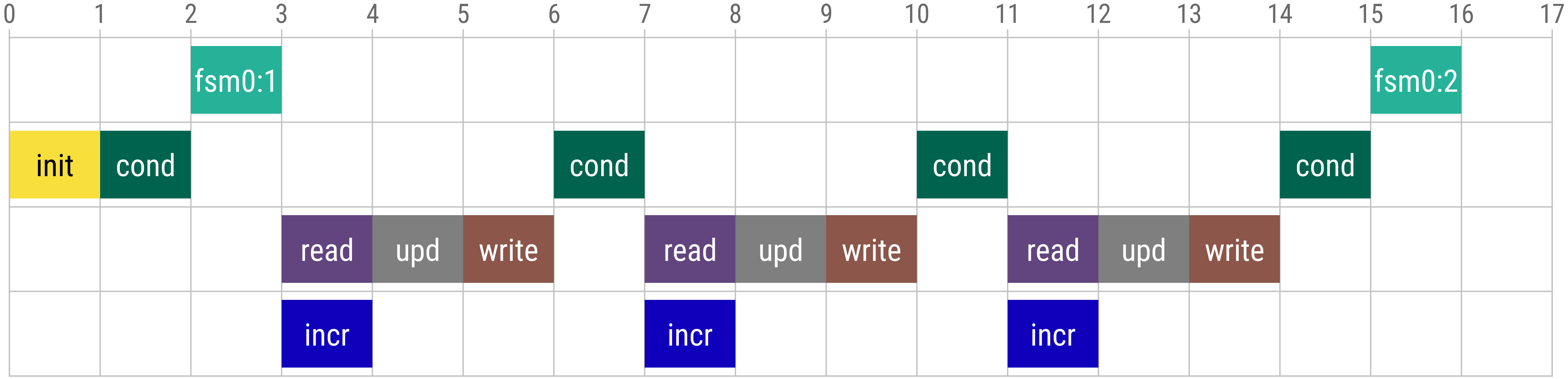}
      \caption{Manually transformed program: Total 16 cycles.}
      \label{fig:control-cycles:while:desugared-timeline}
  \end{subfigure}}

  \makebox[0.45\textwidth][c]{
    \begin{subfigure}{0.49\textwidth}
      \includegraphics[width=\linewidth]{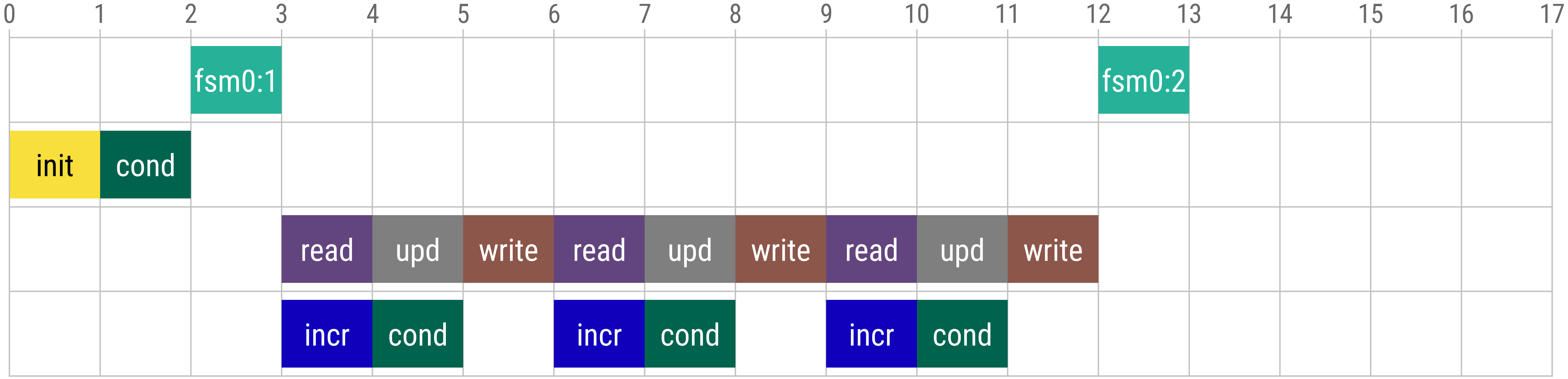}
      \caption{\code{par} optimized program: Total 13 cycles.}
      \label{fig:control-cycles:while:smart-timeline}
  \end{subfigure}}
  \caption{Timelines of three different versions of the example \code{while} program.}
\end{figure}

\paragraphstyle{Setup} The Calyx \code{control} block in
Figure~\ref{fig:control-cycles:while:code:original} updates a computed
number in memory a specified number of times. The \code{while} block
on line~\ref{fig:control-cycles:while:cond-group} drives the
combinational group \code{cond}, which updates the \code{lt} cell's
values based on whether we have reached the loop bound.
\paragraphstyle{Identifying the problem} When we run this program
through the profiler (on inputs where the \code{while} loop runs for
three iterations before terminating), we obtain the timeline on
Figure~\ref{fig:control-cycles:while:original-timeline}.
First, we note that every call of \code{cond} takes one cycle. This
might be insightful for users who expected \code{cond} to be purely
combinational.
Additionally, the timeline shows that there are \emph{three} FSM
register updates (boxes on the top row):~(1) on cycle 2 between
\code{init} and \code{cond},~(2) between \code{cond} and the beginning
of the \code{while} loop, and~(3) after the end of the \code{while}
loop. (2) and (3) are necessary register update cycles because
\code{while} loops cannot be made static due to their fundamental
dynamic nature. However, (1) seems redundant since both \code{init}
and \code{cond} take one cycle each.

\paragraphstyle{Solution} We manually transformed the code
(Figure~\ref{fig:control-cycles:while:code:desugared}) to not use a
combinational component in an attempt to remove (1), which resulted in
a one-cycle cutoff. Additionally, in this example code, we notice that
we could leverage the parallelism at play by performing the counter
check in the same thread as the counter
increment. Figure~\ref{fig:control-cycles:while:code:smart} shows the
result, reducing three more cycles as expected.

%



\subsubsection{Summary}\label{sec:control-cycles:queues:summary}

We apply the three control overhead strategies previously
explored---converting code to static
(Section~\ref{sec:control-cycles:ffnn}); fixing \code{switch-case}
implementation (Sections~\ref{sec:example}
  and~\ref{sec:control-cycles:queues:switch-case}); transforming
  \code{while} loop conditions (Section~\ref{sec:control-cycles:while})--- to the packet scheduling program.
  Table~\ref{fig:control-cycles:queues:opt-strats} shows the effects of
  each optimization strategy.
  We find that each strategy
  contributes at least 10\% of cycle reductions.
  By employing hand-optimization strategies inspired by profiler
  feedback, we were able to reduce a total 46.9\% of
  cycles. Additionally, we used Vivado~\cite{vivado} to obtain
  post-place-and-route resource and critical path estimates for both
  original and optimized versions of the packet scheduling program. We
  found the worst slack slightly increased from 2.263 to 2.499, and
  area \emph{decreased}, where the total number of LUTs went from
  1,251 to 872. We conclude that our final design optimizes both cycle
  counts and area, and has a comparable critical path.

  \begin{table}
    \begin{center}
      \begin{tabular}{lrr}
        \toprule
        \textbf{Strategy} & \textbf{Cycles reduced} & \textbf{\% Cycles reduced} \\
        \midrule
        \texttt{static} & 243,032 & 16.3\% \\
        \texttt{switch-case} & 264,740 & 17.7\% \\
        \texttt{while} & 162,052 & 10.9\% \\
        Other & 28,305 & 1.9 \% \\
        \midrule
        Total & 698,129 & 46.9\% \\
        \bottomrule
      \end{tabular}
    \end{center}
    \caption{Various optimization strategies and the cycle reductions they made on the packet scheduling program.}
    \label{fig:control-cycles:queues:opt-strats}
  \end{table}


  \section{Related Work}\label{sec:related}

  Traditional software profiling and profiling-based performance
  diagnosis~\cite{gprof, hauswirth2005automating, jovic2011catch}
  help software developers identify performance bottlenecks.
  There is also work on profilers for specific software domains, such as high-performance computing~\cite{hpctoolkit2010,
    hollingsworth1991integration,
  kousha2019designing, agelastos2016continuous, sun2021daisen}.
  Kremlin~\cite{Garcia_Kremlin_PLDI_2011}
  identifies regions of serial programs that could most benefit from
  parallelization by using a hierarchical critical path analysis.
  This paper's focus on languages that compile to hardware comes with certain new challenges:
  namely, the added complexity of reconstructing the program's execution state on each cycle,
  and the need to capture fine-grained parallelism in the form of
  call trees (Section~\ref{sec:adl-profiling}).

  Preexisting work on profiling for higher-level accelerator generator
  languages focuses on traditional C++-based high-level synthesis (HLS)
  compilers. HLScope and HLScope+~\cite{choi2017hlscope,
  choi2017hlscope+} instrument users' C++ code with FIFOs and
  dataflow pragmas in order to attribute stalls and perform cycle count
  estimation for loops. \hlsprofiler~\cite{sumeet2022hls_profiler} is a
  framework built on Vitis HLS which performs non-intrusive line-by-line
  profiling of C programs from signals available at C/RTL
  co-simulation. The tool uses FSM values obtained from the synthesis
  Data Analysis Report to map between the RTL simulation trace and the
  source code.
  \xxx[as]{Let's end this paragraph with a one-sentence reason why this existing work does not suffice to solve the
    problems that \profiler solves. One difference is our focus on understanding the compiler, and especially control
    overheads, instead of on understanding line-by-line timing of the source program. Maybe we also need to say why this
  prior work was able to use FSM values instead of instrumentation, where we had to resort to instrumentation?}

  Researchers have also worked on HLS-based profiling that targets
  \emph{in-FPGA execution}, as opposed to software
  simulation. HLS\_Print~\cite{sumeet2021hls_print} is a logging
  framework that runs on an FPGA in parallel with the hardware design,
  but it is intrusive in nature.
  \xxx[as]{What does ``intrusive in nature'' mean? Are we \emph{not} intrusive?}
  Curreri et al.~\cite{curreri2010}
  created a runtime performance analysis of Impulse C HLS programs, but
  it uses FSM values to map back to the HLS source code
  level. RealProbe~\cite{realprobe} instruments
  user HLS code to capture control signals (analogous to \code{go}/\code{done} signals
  used by Calyx) to track C functions and loops, and records results to
  DRAM for more accurate cycle-count information than HLS
  simulation. While \profiler does not perform in-FPGA profiling, we
  perform profiling at a more fine-grained level.
  \xxx[as]{Can we get more specific? What exactly can we ``see'' that these coarser-grained approaches cannot?}
  Furthermore, this
  fine-grainedness enables us to separate cycles devoted to actual
  computation as opposed to scheduling updates.
  \xxx[as]{This could be phrased as ``measuring control overhead,'' to match with terminology used elsewhere.}


  Profilers that focus on commercial C/C++ HLS compilers restrict the input languages they can support
  and rely on closed-source commercial infrastructure.
  On an engineering level, \profiler targets the open-source Calyx intermediate language and
  supports any ADL
  that compiles to Calyx.

  Waveform visualization tools~\cite{gtkwave, surfer} help users
  navigate RTL simulation traces, but they only reflect RTL-level
  signals which are difficult to trace back to the ADL
  level. Cider~\cite{cider} is a debugger and interpreter for Calyx that
  aids ADL users in finding correctness bugs in their code.
  \profiler can complement Cider by helping ADL users identify
  performance bugs. 

  \section{Conclusion}\label{sec:conclusion}

  ADL compilers exhibit a necessary unpredictability because of the
  complexity of converting software idioms to hardware structure. By
  creating \profiler and using it to find optimization opportunities
  missed by the Calyx compiler, we advocate for the necessity of
  compiler understanding tools. Such tools can help ADL users navigate
  choices made by compilers, work around cases where the compiler
  makes sub-optimal scheduling choices, and make ADLs even more
  accessible and realistic for user bases.

  \bibliographystyle{./acm/ACM-Reference-Format}

  \interlinepenalty=10000

  \bibliography{./bib/venues,./bib/papers}

  \end{document}

%% file: figures/switch-case-code.tex
\begin{figure}
\begin{lstlisting}[numbers=left, belowskip=-0.8\baselineskip, escapechar=|]
component main() -> () {
  cells {
    @external mem = comb_mem_d1(32, 1, 1);
    r = std_reg(32);
    ans = std_reg(32);
    s1 = sub_one();
    s2 = sub_two();
    s3 = sub_three();
    eq_1 = std_eq(32);
    eq_2 = std_eq(32);
    eq_3 = std_eq(32);
  }
  wires {
    group read {|\label{switch-case:group-read}| // read mem into register r
      mem.addr0 = 1'd0;
      r.in = mem.read_data;
      r.write_en = 1'd1;
      read[done] = r.done;|\label{switch-case:group-read:done}|
    }
    group run_s1 {|\label{switch-case:group-invoke-one}|
      s1.go = 1'd1;
      run_s1[done] = s1.done;
    }
    ... // groups run_s2, run_s3, write
    eq_1.left = r.out;|\label{switch-case:continuous}|
    eq_1.right = 32'd1;
    ... } // similar assignments for eq_2, eq_3
  control {|\label{switch-case:control}|
    seq { |\label{switch-case:seq-start}|
      read; |\label{switch-case:read-enable}|
      par { |\label{switch-case:par-start}|
        if eq_1.out { run_s1; } |\label{switch-case:if}|
        if eq_2.out { run_s2; } |\label{switch-case:case2}|
        if eq_3.out { run_s3; } |\label{switch-case:case3}|
      }|\label{switch-case:par-end}|
      write; } } }
\end{lstlisting}
\caption{A Calyx component generated from Figure~\ref{fig:switch-case-adl}. The user-defined components \code{sub_one}, \code{sub_two}, and \code{sub_three} are omitted. Each \code{case} is checked in parallel.}
\label{fig:switch-case-calyx}
\end{figure}